# Corrosion resistance of a water-borne resin doped with graphene derivatives applied on galvanized steel

A. Collazo, B. Díaz, R. Figueroa, X.R. Nóvoa, C. Pérez *

*CINTECX, Universidade de Vigo, ENCOMAT Group, 36310 Vigo, Spain*

ABSTRACT

The present work reports the effect of adding Graphene Oxide (GO) and reduced Graphene Oxide (rGO) in the corrosion protection provided by a water-borne resin applied on a galvanized steel substrate. Three concentrations, 0.05, 0.1 and 0.15 (all wt%) were tested. The results were markedly affected not only by the concentration of particles but also by their nature. Although the zeta potential values suggested good dispersibility of the particles in the resin, certain aggregation was observed, mainly in rGO 0.1 wt% and rGO 0.15 wt% formulations. The electrochemical impedance spectroscopy (EIS) technique characterised the free films' transport properties. The results suggested that the aggregation strongly influenced the film morphology. The rGO 0.1 wt% and rGO 0.15 wt% formulations exhibited percolating pores that facilitated the electrolyte uptake through the films. The EIS technique was also used to study the protective performance of the films when applied to the metallic substrate. The results confirmed the harmful effect of the particle's aggregation. The results were interesting for the rGO 0.05 wt% system, which displayed long-lasting protection properties. This performance was explained considering its good barrier properties and the zinc surface passivation by the generation of zincite, ZnO.

## 1. Introduction

The application of organic coatings is a common approach for protecting of metallic structures exposed to aggressive environments. The polymer film is supposed to act as a barrier that isolates the metal from the corrosive surroundings [1]. This barrier effect can be enhanced by using lamellar pigments that delay the water, oxygen, and ions flow toward the metallic substrate. Formulations that include fillers such as lamellar zinc phosphate [2], mica [3], layered double hydroxides (LDHs) [4], or natural clays [5] are examples of improving the performance of organic coatings. However, their efficiency is limited to a short exposure period. For this, the investigation is still going on for more efficient and long-lasting alternative materials. In recent years, several research outcomes suggest that graphene derivatives, namely graphene oxide (GO) and reduced graphene oxide (rGO), can be good candidates for manufacturing protective layers [6]. Both species exhibit advantageous features, such as the two-dimensional morphology, with a very high aspect ratio. Also, they are mechanically strong and chemically inert [7]. However, both exhibit important structural differences; consequently, their chemical properties are significantly distinct.

Graphene oxide (GO) is considered the oxidized form of graphene.

During the oxidation process, the oxygen-containing groups such as hydroxyl or epoxy are incorporated into the basal plane, while carboxyl or carbonyl groups are located at the periphery. The presence of these oxygen functional groups explains its hydrophilic character, which improves its chemical reactivity and dispersibility in water or other polar solvents, even though these characteristics strongly depend on the C/O ratio, i.e., the oxidation degree. One important advantage of the GO is its compatibility with a variety of polymeric matrixes, mainly water-borne resins, which contributes to developing more eco-friendly formulations. The GO reduction leads to the reduced graphene oxide (rGO), a compound with a graphene-like structure, although it still contains residual oxygen and structural defects that originated during the GO synthesis. The removal of the oxygen groups markedly increases the C/O ratio. The associated structural transformations lead to drastic changes in the rGO properties. Especially noteworthy is the large increase in the electrical conductivity due to the partial regeneration of the π-conjugated system. On the other hand, the lack of these electronegative groups favours the hydrophobicity character and the chemical inertness; these features make it difficult to obtain stable and uniform dispersions, which is essential to provide good barrier performance [7–9].

The surface modification of GO or rGO is a common practice to improve the dispersion in polymers. Representative examples of this functionalization are the use of silane coupling agents [10,11], amines [12,13], polyaniline (PANI) [14], or surface grafting with nanoparticles (silica, titanium oxide, or alumina are the most studied) [15–17] among others. Generally, the functionalization improves the dispersibility, but it also involves a certain breakage of the aromatic structure, which can lead to poorer barrier behaviour, mainly in the case of rGO [6].

* Corresponding author.
*E-mail address:* cperez@uvigo.es (C. Pérez).

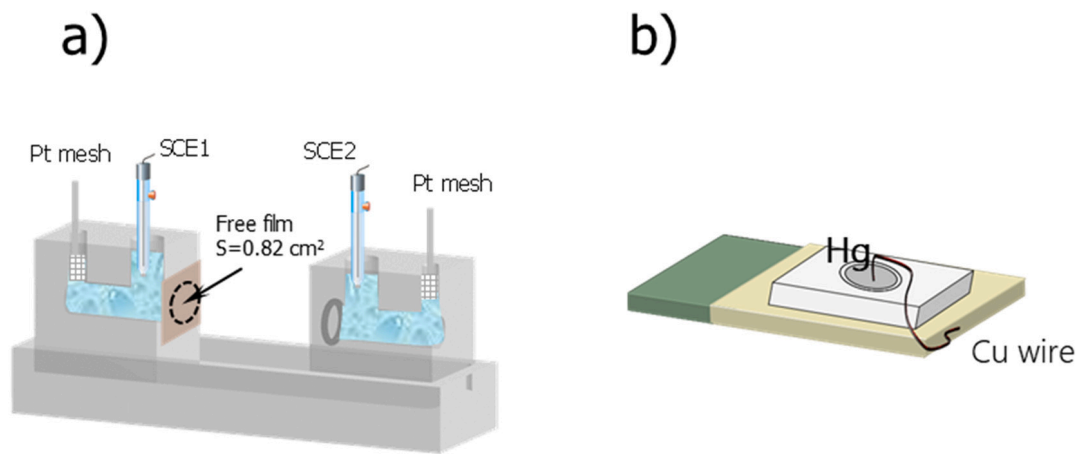

**Fig. 1.** Sketch of the permeation cell used to perform the transport properties of the free films (a). Drawn of the mercury pool used to characterize the dry film properties (b).

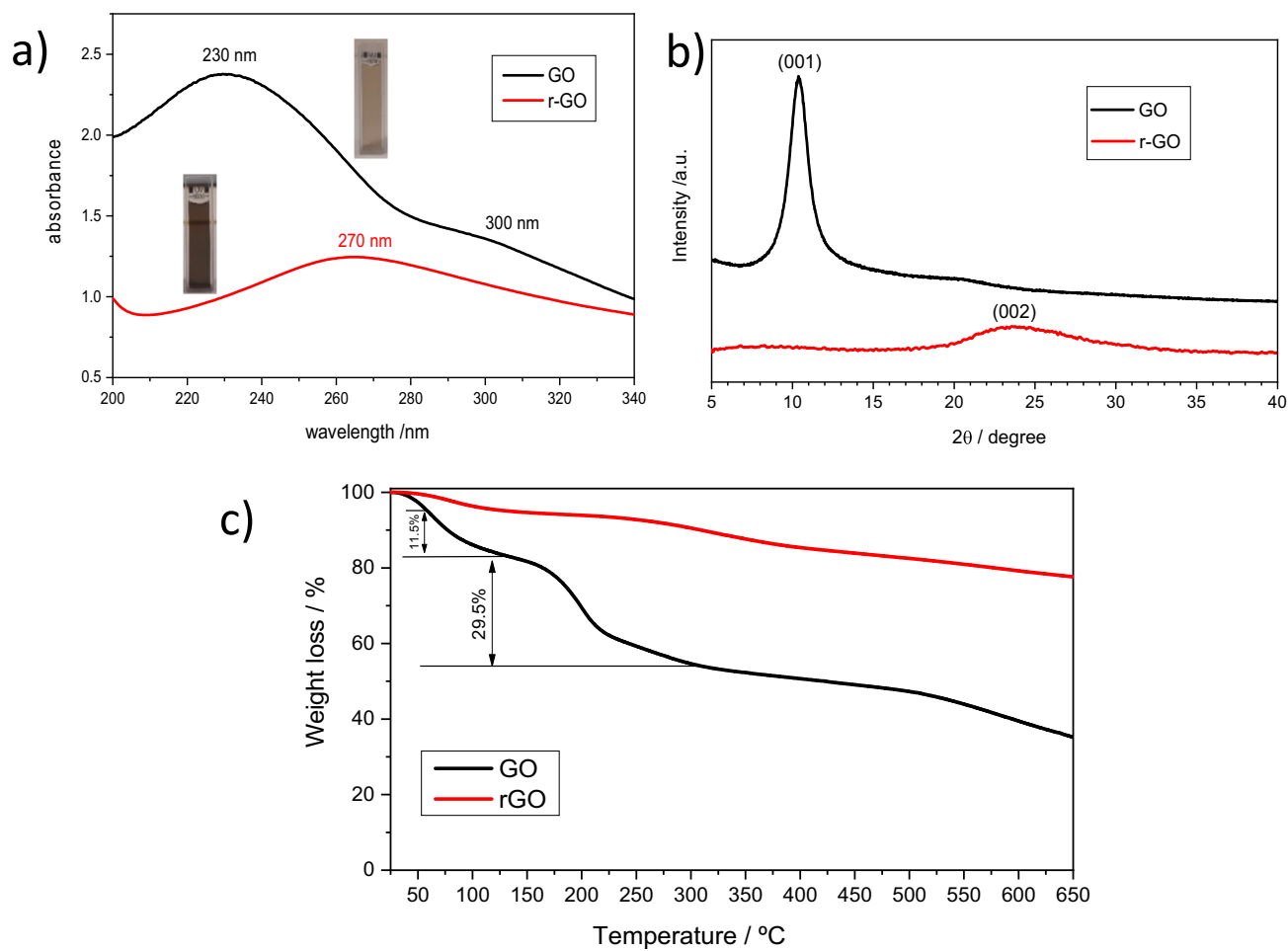

**Fig. 2.** UV–vis spectra for GO and rGO aqueous dispersions. Photographs of the suspensions are inset (a). XRD spectra (b) and thermogravimetric plots of GO and rGO (c).

Additionally, these chemical modifications reduce the electrical conductivity of the rGO particles, an aspect of great importance when the anticorrosive coating is intended to provide cathodic protection. In this line, some studies dealt with the addition of conductive fillers, such as graphene derivatives, to zinc-rich paints that would reduce zinc particles' content while maintaining the electrical contact between the Zn particles and with the metal substrate [18].

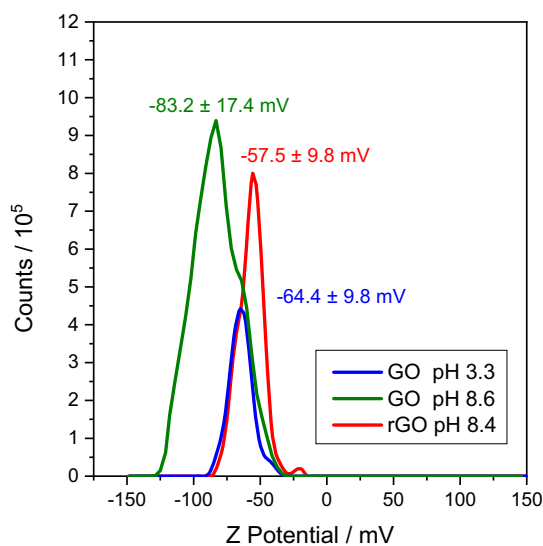

**Fig. 3.** Zeta potential plots of GO and rGO aqueous dispersions at different values of pH.

All those approaches have been of limited success, mainly in long-term protection. In addition, the experimental procedures are usually complex, which restricts the implementation on an industrial scale. In the present study, GO, or rGO aqueous suspensions were directly mixed with an acrylic water-borne resin by mechanical and sonication procedures [19]. Different GO/rGO concentrations were tested to assess the amount that provides the best anticorrosive behaviour when applied on a galvanized steel substrate. On the other hand, as the addition of particles usually modifies the coating morphology, mainly the porosity, we have characterised the free films' transport properties to quantify such effect.

## 2. Experimental

### 2.1. Materials

The water-borne resin used was prepared ad hoc for the present study by Akzo Nobel Coatings; it is based on an acrylic resin with a total solids content of 35.76 % and a density of 1.024 g mL$^{-1}$. The employed GO was a commercial water dispersion of graphene oxide sheets from Graphenea® with 0.4 wt% concentration. The particle size determined by SEM technique was less than 10 μm in the largest dimension (this data was provided by the manufacturer). The rGO was obtained by chemical reduction of dispersed GO sheets by reaction with hydrazine as a reducing agent [20]. Briefly, 75 mL of deionized water was mixed with 25 mL of the GO suspension (0.4 wt%) and 130 μL of hydrazine monohydrate. Three drops of 25 % ammonia solution were also added to adjust the pH to 10, to promote the colloid stability. The reaction was carried out in a water bath at 95 °C for 4 h. The residue was washed several times with deionized water and dried at 30 °C for about 24 h.

Three different GO/rGO concentrations were tested: 0.05 wt%, 0.1 wt%, and 0.15 wt% with respect to the total solid content. The mixtures were mechanically stirred for 30 min; after that period, uniform colour was observed. The formulations were applied by brush on galvanized steel plates with 100mmx50mmx2mm dimensions and a nominal zinc thickness of 20 μm on each side. The films were cured for 24 h in ambient laboratory conditions. The resulting average dry film thickness was 77 μm ± 8 μm.

Some experiments upon free films were performed to assess the effect of the GO/rGO additions on the transport properties of the resin. For that, the different formulations were applied on a Poly-methyl methacrylate (PMMA) sheet and, after curing, they were carefully detached.

### 2.2. Experimental techniques

The GO and rGO characterization was performed by the following techniques:

– UV–vis absorption spectroscopy, using a double-beam spectrophotometer Specord® 200 Plus within a scan range of 200–350 nm at a scan rate of 10 nm s$^{-1}$.
– X-Ray Diffraction (XRD) technique with a Siemens® D5000 powder diffractometer with the monochromatic CuK$_\alpha$ radiation with λ = 1.5 Å, the step size and time were 0.02° and 4 s, respectively.
– Thermogravimetric analysis (TGA), using a Mettler Toledo® TGA/DSC 1 instrument, from 23 °C to 650 °C, at a heating rate of 10 °C min$^{-1}$, in an N$_2$ stream flow of 50 mL min$^{-1}$.
– Zeta potential measurements, made in a Malvern Zetasizer NanoZS® instrument. The reproducibility was verified by performing a minimum of 10 measurements per set. The zeta potential was calculated by determining the electrophoretic mobility using the Smolukowski model.

The surface morphology of the coated samples was assessed by collecting three-dimensional images using a Dektak 150 Surface Profiler® (Veeco). Additionally, a digital microscope Dino-Lite AM9515MZT-EDGE® was used to characterize the particles' dispersion in the film once cured. The morphological characterization was carried out by Scanning Electron Microscopy (SEM) and Energy dispersive X-ray (EDX) techniques. The equipment employed was an Electroscan JSM-54 model JEOL® 5410 equipped with an energy dispersive X-ray detector Link ISIS® 300. The identification of the crystalline corrosion products was performed using the X-ray diffractometer afore-described.

The transport properties of the free films were assessed by EIS measurements using a permeation cell, as displayed in Fig. 1a. Both compartments were filled with 0.1 M KCl, and two O-rings define a film surface of 0.82 cm$^2$ in contact with the electrolyte on both sides [21].

A four-electrode arrangement was used with two external Pt mesh as current-driven electrodes and two saturated calomel electrodes (SCE) placed close to the sample, acting as sensing electrodes. The EIS measurements were made with an Autolab 30 potentiostat from Ecochemie®, at null d.c.c, and a 10 mV$_{rms}$ sinusoidal perturbation was applied. The frequency range was swept from 100 kHz to 1 Hz, with seven frequency points per decade.

The protective properties of the different formulations were also evaluated by EIS measurements using the aforementioned equipment and a three-electrode arrangement. The coated galvanized steel, with an 8 cm$^2$ nominal surface, was the working electrode; an SCE and a Pt mesh were used as the reference and counter electrodes, respectively. The electrolyte was 0.1 M NaCl solution. The equipment and the set of parameters were the same as used to perform the free film EIS measurements, although the low-frequency limit was 10 mHz.

High-frequency EIS experiments were performed for the characterization of the coated samples in dry conditions using an Agilent 4294A® impedance-phase analyser. The employed cell was a mercury pool, which delimits a working surface of 0.78 cm$^2$; Fig. 1b shows a schematic view of the arrangement. The scanned frequency range was from 1 MHz to 1 kHz in pure AC mode, and 15 data points per decade were collected.

At least three samples were tested for each system to verify the reproducibility of the obtained results.

## 3. Results and discussion

### 3.1. Characterization of the GO and rGO employed

The chemical reduction of GO was corroborated by several analytical techniques. Fig. 2 depicts the UV–vis spectra of the GO and rGO. The spectrum obtained for GO exhibits a characteristic peak around 230 nm associated with the π–π* of the aromatic C—C bonds, and a shoulder at

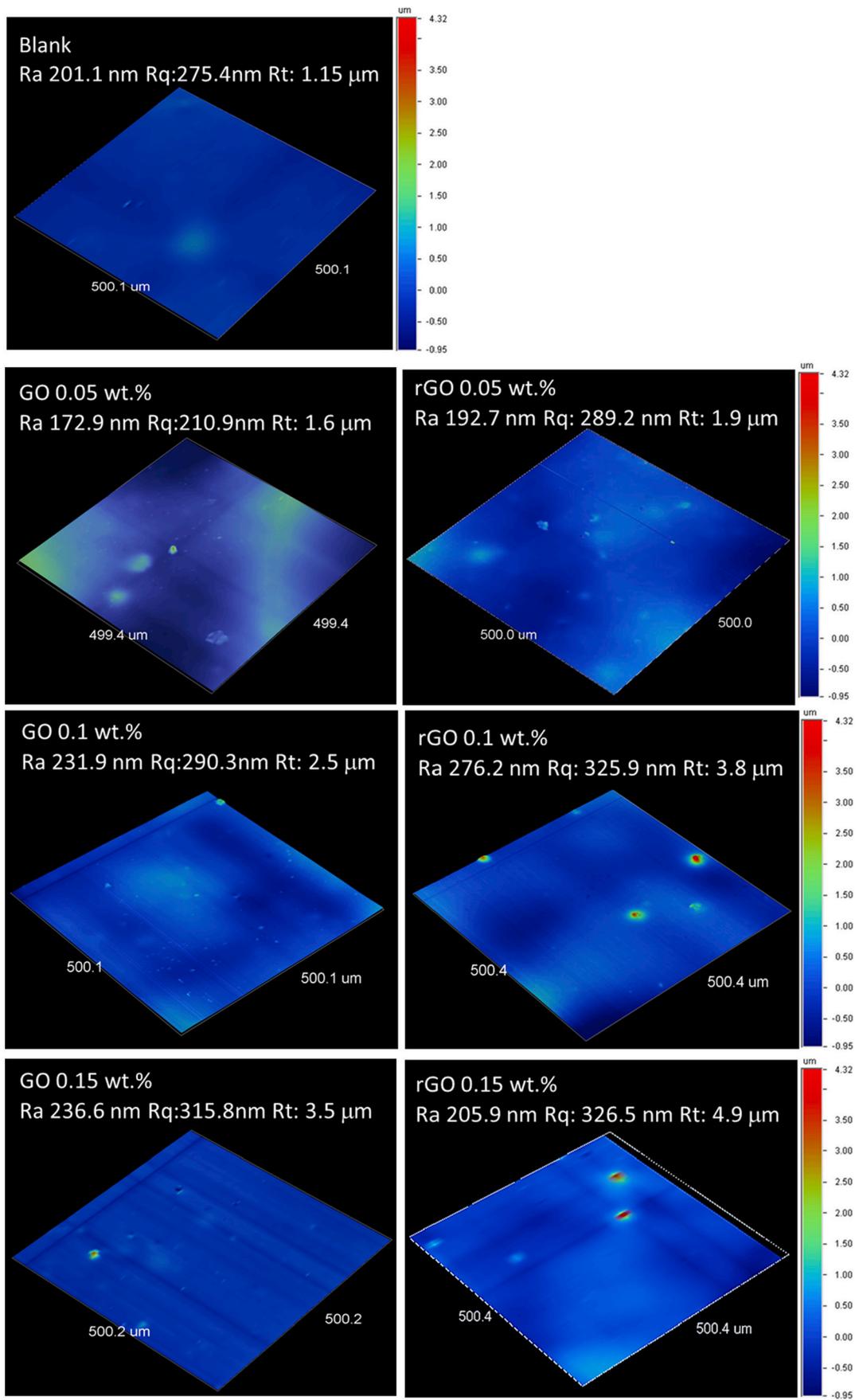

**Fig. 4.** Three-dimensional images of the galvanized steel samples coated with the different formulations tested. The roughness parameters, Ra, Rq, and Rt, are included.

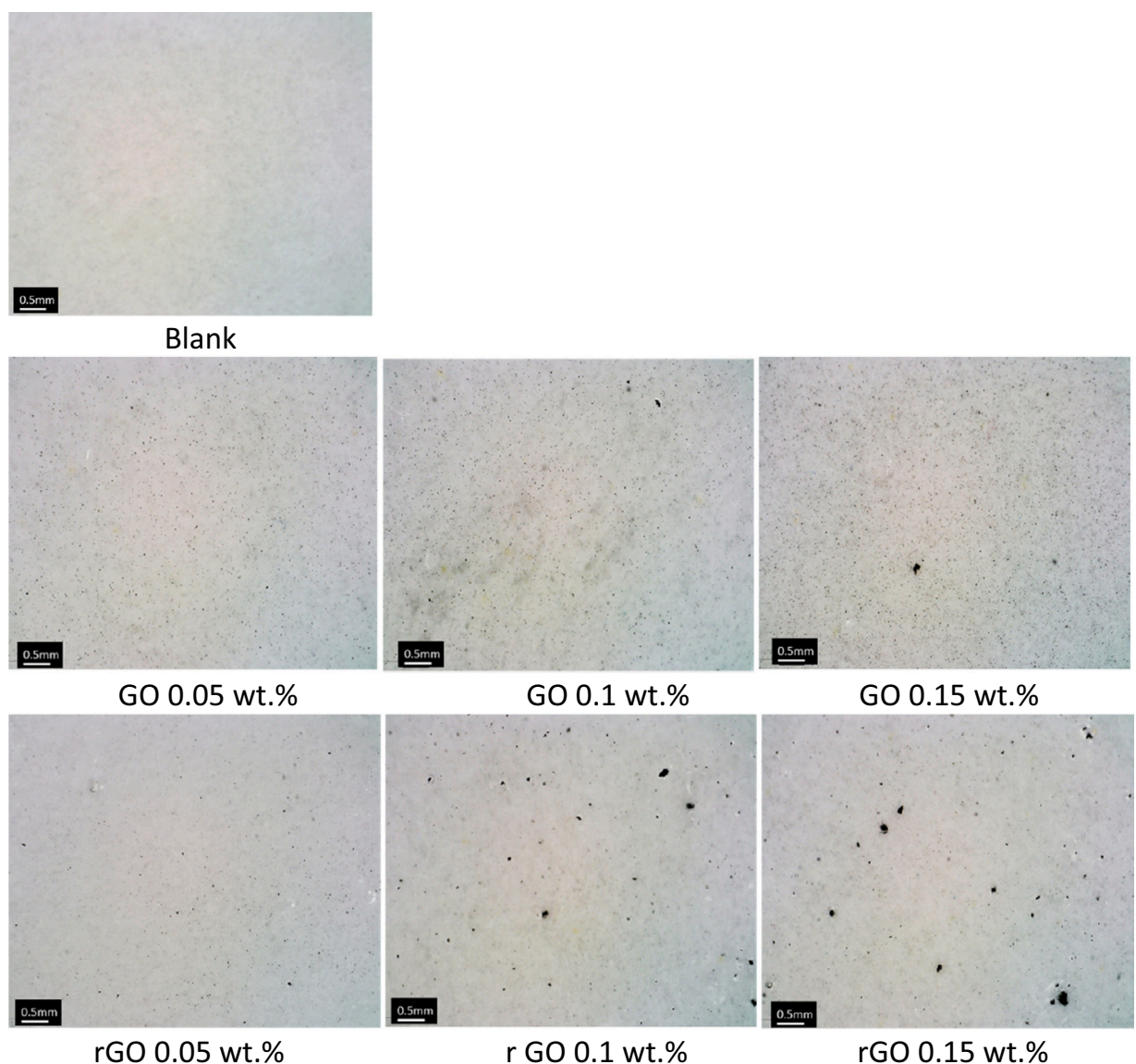

**Fig. 5.** Pictures of the films applied on a clear glass to observed the particles' distribution.

300 nm ascribed to the n-π* of C=O bonds. The characteristic peak observed in the rGO spectrum is red-shifted to about 270 nm, indicating that the electronic conjugation was restored [20,22]. Pictures of both aqueous suspensions are shown inset in Fig. 2a. A significant colour change from brown (GO) to black (rGO) can be appreciated, which corroborates the electronic restoration in the rGO.

The XRD patterns of GO and rGO are displayed in Fig. 2b. The GO spectrum presents the (001) diffraction peak at 2θ = 10.5°, which corresponds to a layer-to-layer distance (d-spacing) of 0.84 nm. This interlayer expansion is a consequence of the oxygen-containing functional groups attached to the GO sheets. Instead, the rGO pattern exhibits a broad (002) diffraction peak around 2θ = 23.5° with a lower d-spacing (0.38 nm), due to the restacked graphene particles when the oxygen groups are removed [22,23].

The thermal stability of GO and rGO was investigated by TGA. The obtained TG plots are presented in Fig. 2c. The GO exhibits an initial weight loss at temperatures lower than 100 °C corresponding to the removal of the adsorbed water. However, the main mass loss takes place at about 200 °C (∼ 30 %), which is attributed to the decomposition of the labile oxygen functional groups, whereas the steady mass loss (∼23 %) observed at temperatures higher than 300 °C is ascribed to the released of the more stable oxygen functional groups. In contrast, the rGO does not exhibit the abrupt mass loss at 200 °C which reveals the efficiency of hydrazine in removing the labile oxygen functionalities. However, there is still a steady mass loss at temperatures above 100 °C (∼18 %) suggesting that the more stable oxygen groups were not eliminated in the rGO. These results are in agreement other research [20,22,24,25].

Assessing the stability of the GO/rGO aqueous dispersions is a critical aspect to design new formulations based on water-borne resins. The zeta potential determination is a common technique for characterizing the stability of colloidal dispersions. It provides a measure of the magnitude and sign of the surface charge associated with the double layer around the colloid particle [26]. The zeta potential measurements were performed at the natural pH of the corresponding dispersions, which were 3.3 for GO and 8.3 for rGO. The pH of the acrylic water-borne resin was around 8.5 (this information was provided by the manufacturer). Therefore, zeta potential measurements on GO suspensions at this pH were also performed. The results are displayed in Fig. 3.

It is widely accepted that absolute values higher than 30 mV lead to stable dispersions [26]. As it can be seen in Fig. 3, the zeta potential values measured are well below this threshold, which supports the viability to obtain good dispersibility of these particles into the resin. The GO/rGO aqueous dispersions have negative zeta potential values, as a consequence of the ionization of the oxygen functional groups, which is more pronounced at higher pH. For similar pH, the higher absolute zeta potential values correspond to GO where the density of oxygen functional groups is higher. These results are in agreement with those reported in the literature [23,26,27].

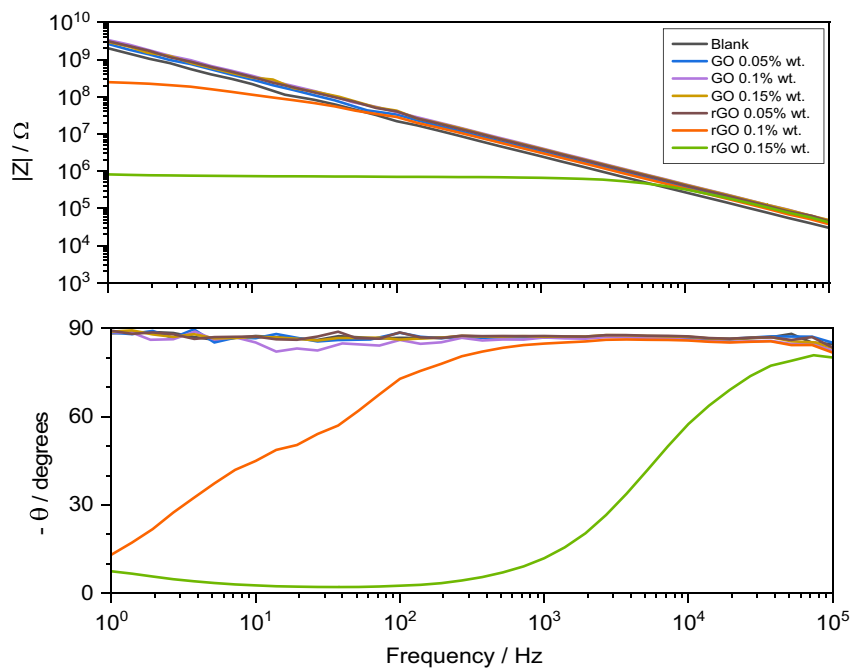

**Fig. 6.** Bode plots corresponding to the free films of the studied formulations after 6 days in contact to the 0.1 M KCl solution.

The surface morphology of the applied films was studied by mechanical profilometry. Fig. 4 shows the 3D images obtained for the different formulations following the ISO 4288 procedure [28]. Besides, the values of some standardised roughness parameters frequently used to characterize the surface morphology (Ra, Rq, and Rt) are included. The roughness average, Ra; and RMS roughness, Rq, defined respectively as the arithmetic average of the absolute values of the profile heights and the roof root mean square average of the profile heights [29], tend to increase with the content of the particles, as it was expected. The Rt parameter, corresponding to the vertical distance between the highest and lowest points of the profile [29], undergoes a remarkable increase in the rGO 1 wt%, rGO 0.15 wt%, and GO 0.15 wt% formulations, with values higher than 3.5 μm. Especially higher is in systems doped with rGO (0.1 and 0.15 wt%) that supports the aggregation tendency of these particles. However, at low concentrations (0.05 wt%) the dispersion seems to be adequate according to the Rt values, similar to those obtained for GO 0.05 wt%.

The aggregation tendency was corroborated by visual inspection of the coatings applied on a glass substrate to better visualize the particles' distribution. The results are displayed in Fig. 5. The particles are distinguished in all the doped films, although the aggregates are appreciated in the more concentrated formulations. Highlights the uniform particles' distribution in samples doped with GO, only certain agglomeration is observed in GO 0.15 wt% samples. It is also observed homogenous particles' distribution at low rGO concentration (0.05 wt%), however significant aggregation is appreciated at rGO 0.1 wt% and rGO 0.15 wt% concentrations.

### 3.2. Study of transport properties of the free films

Free film experiments were performed to characterize the transport properties of the coatings without the influence of the metallic substrate.

Fig. 6 displays the characteristic Bode plots recorded for the different free films in contact with a 0.1 M KCl solution.

Most of the systems present a capacitive behaviour characterised by a straight line in the impedance modulus and a phase angle close to 90°. Thereby, good barrier properties can be expected. However, when the resin is doped with rGO 0.1 wt% and 0.15 wt%, the impedance values are markedly lower, mainly for the rGO 0.15 wt% system, and a low-frequency limit is clearly observed. These results indicate the presence of percolating pores, possibly due to the particles' aggregation stated by the profilometry experiments. The higher particle size should negatively affect the polymer crosslinking.

The coating capacitance, and so the transport properties of the films, may be extracted from the complex capacitance plots (also named Cole-Cole capacitance) by extrapolation to infinite frequency. The capacitance, $C(\omega)$, can be calculated from the experimental impedance data, $Z(\omega)$, considering:

$$Z(\omega) = R_e + \frac{1}{j\omega C(\omega)} \ and \ so \ C(\omega) = \frac{1}{j\omega(Z(\omega) - R_e)} \tag{1}$$

The real and imaginary parts of the capacitance are:

$$C'(\omega) = \frac{-Z''(\omega)}{\omega \left[ (Z(\omega) - R_e)^2 + Z''(\omega)^2 \right]} \tag{2}$$

$$C''(\omega) = -\frac{Z'(\omega) - R_e}{\omega \left[ (Z(\omega) - R_e)^2 + Z''(\omega)^2 \right]} \tag{3}$$

where $R_e$ accounts for the electrolyte resistance, $\omega$ corresponds to the angular frequency, and $Z'(\omega)$, $Z''(\omega)$, $C'(\omega)$, and $C''(\omega)$ are the real and imaginary parts of the impedance and capacitance, respectively [30,31].

Once the coating capacitance is extracted, the complex dielectric constant is obtained considering a plane condenser behaviour (Eq. (4)):

$$C(\omega) = \varepsilon(\omega)\varepsilon_o \frac{A}{d} \tag{4}$$

where $\varepsilon(\omega)$ and $\varepsilon_o$, represent the complex dielectric permittivity ($\varepsilon(\omega) = \varepsilon' + j\varepsilon''$) and the vacuum permittivity, respectively. A is the surface area in contact with the electrolyte and d is the film thickness.

The permittivity was calculated at the highest measured frequency

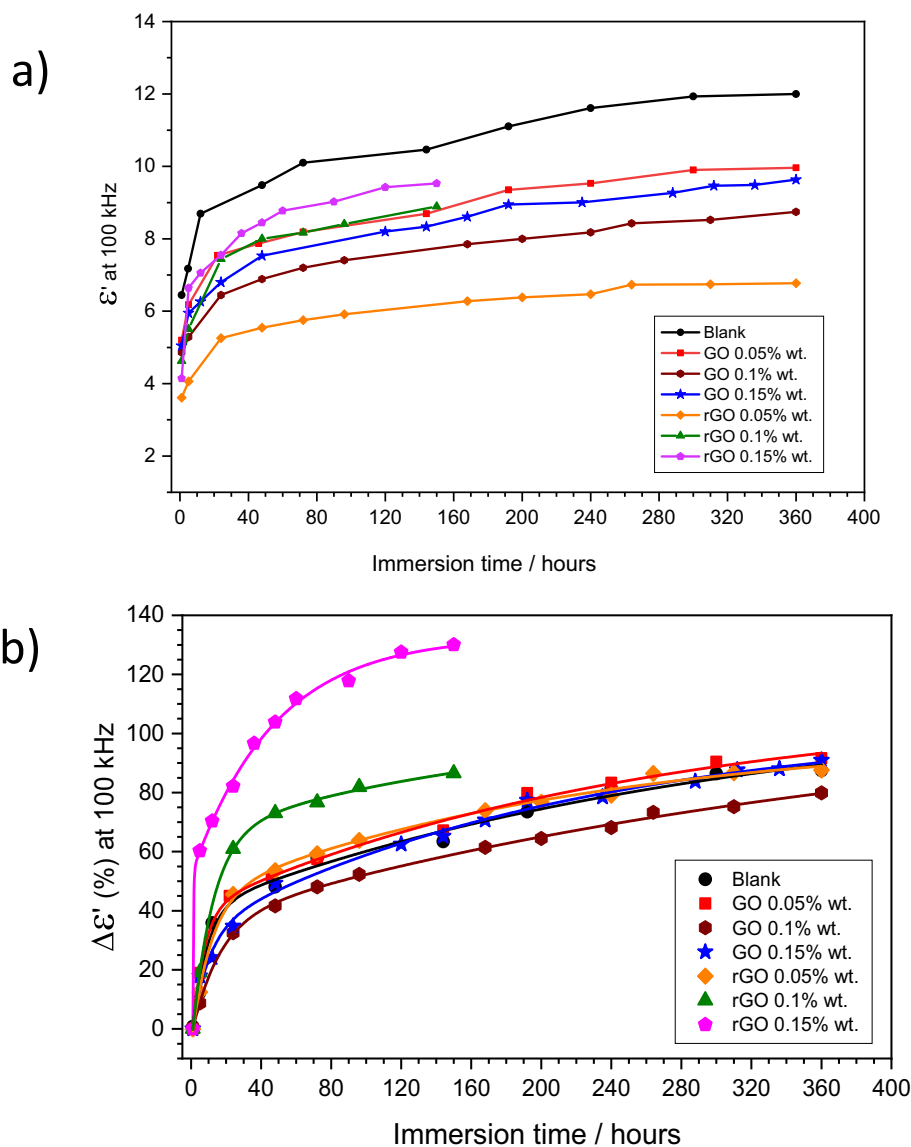

**Fig. 7.** Evolution of the real part of the permittivity (ε') (a) and evolution of its relative variation (Δε') (b) with immersion time in 0.1 M KCl solution.

**Table 1**
Parameters obtained from the non-linear fitting of the data of Fig. 7b using Eq. (5).

| System | $y_o$ | $A_1$ | $\tau_1$ (h) | $A_2$ | $\tau_2$ (h) |
|---|---|---|---|---|---|
| Blank | 110.8 | 46.6 | 7.9 | 70.6 | 300.8 |
| GO 0.05 wt% | 110.2 | 46.9 | 6.6 | 70.3 | 252.7 |
| GO 0.1 wt% | 109.6 | 38.7 | 15.5 | 73.6 | 399.2 |
| GO 0.15 wt% | 103.6 | 34.2 | 9.8 | 71.1 | 216.3 |
| rGO 0.05 wt% | 99.6 | 49.3 | 11.2 | 55.6 | 219.0 |
| rGO 0.1 wt%. | 102.3 | 71.4 | 11.8 | 37.1 | 173.2 |
| rGO 0.15 wt%. | 132.9 | 962.2 | 0.35 | 81.1 | 47.3 |

(100 kHz) [32] and $R_e$ corresponds to the high frequency limit in the impedance plots.

Fig. 7a depicts the evolution of the real part of the permittivity (ε') extracted from the Cole-Cole capacitance as a function of the immersion time in 0.1 M KCl solution.

All the films track the same pattern; a quick initial increase followed by a quasi-stabilization at a longer immersion period. This evolution was also observed by P. Bonin et al. [33] and it should be attributed to the water uptake during the immersion time. For comparative purposes, the analysis of the relative variation of ε' (Δε') is more useful, since it does not depend on the initial permittivity value. This evolution is shown in Fig. 7b.

As expected, the highest ε' increase corresponds to free films doped with rGO 0.15 wt% followed by rGO 0.1 wt%, in agreement with the presence of percolating pores observed in these two systems. In fact, for these formulations, the immersion period considered was shorter because of the marked impedance decrease observed with respect to the other systems from the beginning (see Fig. 6). The continuous line in Fig. 6b corresponds to the fitting of the experimental data using the Eq. (5):

$$y = y_o - A_1 e^{-t/\tau_1} - A_2 e^{-t/\tau_2} \tag{5}$$

In Eq. (5) $y_o$ represents the asymptotic maximum value of the Δε', and the other two terms correspond to two time constants ($\tau_1$ and $\tau_2$) associated with two different processes related to the water uptake [34,35]. The contribution of each is given by the corresponding $A_i$ values. Table 1 summarized the best fitting values for $y_o$, $A_i$, and $\tau_i$.

Regardless of the system, two well-differentiated kinetics are appreciated, in which the time constants are separated more than one order of magnitude in the time scale. The initial fast process ($\tau_1$) is characterised by a sharp $\Delta\varepsilon'$ increase, probably associated with the filling of surface pores or the more accessible area of the films. The second process presents a slower kinetics ($\tau_2$), and can be due to the water interaction in the bulk of the film and, in the case of films with rGO 0.1 wt% and rGO 0.15 wt%, the diffusion through the percolating pores [34,35]. In addition, it is noteworthy the low values observed for both time constants in the case of films doped with the highest amount of rGO, which supports the hypothesis that the aggregation hinders the adequate network formation. Therefore, bad barrier properties should be expected.

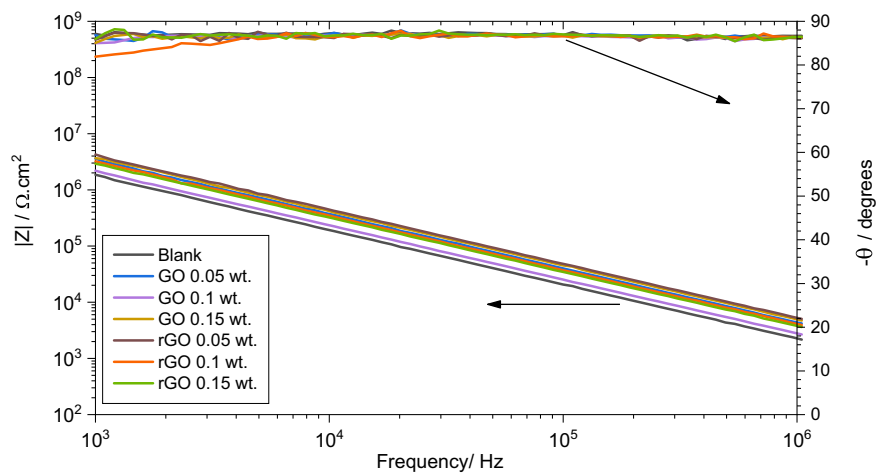

**Fig. 8.** Bode plots obtained from the galvanized steel coated with the different systems in dry conditions.

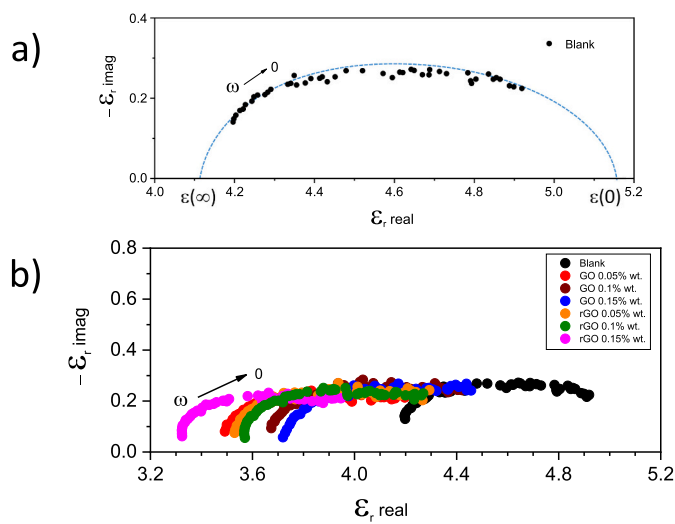

**Fig. 9.** Frequency distribution of the complex relative permittivity of the undoped coating illustrating the $\varepsilon(\infty)$ and $\varepsilon(o)$ meaning (a), and the complex relative permittivity representation for all the systems (b).

### 3.3. Behaviour of the coatings applied on galvanized steel in dry conditions

The previous section analysed the transport properties of the films and how they can be affected by the presence of the particles. This section focuses on the effect of these additions when the systems are applied on a metallic substrate.

Prior to assessing the anticorrosive properties of the films, the characterization of the applied films in dry conditions was performed. For that, high-frequency measurements were made using the experimental setup shown in Fig. 1b. The Bode plots obtained are shown in Fig. 8.

All the formulations exhibit a capacitive behaviour with phase angles around 87° and no low-frequency limit is observed in the scanned frequency range. Since these measurements were performed in dry conditions, the impedance spectra should be related to the relaxation processes of the polymer and how they can be affected by the presence of the particles.

The complex dielectric permittivity was obtained using the procedure above described (Eqs. (1)–(4)). Fig. 9 displays the Nyquist plots of the complex permittivity of the studied systems. A detailed description of dipolar relaxation in polymers and its relation with the EIS measurements is beyond the scope of the present work. Nevertheless, some comments are required prior to discussing the obtained results [36].

The basic model for describing the dielectric relaxation is the Debye-type expression (a single relaxation phenomenon):

$$\varepsilon = \varepsilon_\infty + \frac{\varepsilon_o - \varepsilon_\infty}{1 + j\omega\tau} \tag{6}$$

where $\varepsilon(\infty)$ and $\varepsilon(o)$ represent the high and low-frequency permittivity values, and $\tau$ the dipolar relaxation time constant. This expression corresponds to a semicircle in a complex plane plot, in which the maximum is given by $\omega\tau = 1$. However, most of the real situations deviate from this ideal model and subsequent modifications that take into account the distribution of the time constant have been developed. One of the most popular is the Cole-Cole type distribution:

$$\varepsilon = \varepsilon_\infty + \frac{\varepsilon_o - \varepsilon_\infty}{1 + (j\omega\tau)^\alpha} \tag{7}$$

In this expression $\alpha$ is the parameter that describes the width of the time constant distribution, and its representation in the Nyquist plot exhibits a depressed semicircle.

Fig. 9a) shows an example of the acquired Nyquist plot of the $\varepsilon(\omega)$. The depressed capacitive loop characteristic of the Cole-Cole type distribution can be appreciated. This result reveals the heterogeneous nature of the network generated during the polymer crosslinking process, and thus, the presence of defective areas is probable. The effect of the addition of particles is illustrated in Fig. 9b, all the plots exhibit a Cole-Cole type distribution, analogous to that observed for the undoped film. In addition, regardless of the amount or nature of the particles, a clear decrease in the permittivity value is observed, which should be related to an increase in porosity. It seems that the particles hinder the formation of a continuous network [37]. As a general tendency, it can also be observed that the decrease is stronger when the film is doped with rGO, mainly for rGO 0.15 wt% formulation. This is due to the conducting character of rGO particles. M. Keddam et al. [38] demonstrate that the addition of small amounts of conducting particles to a polymer decreases

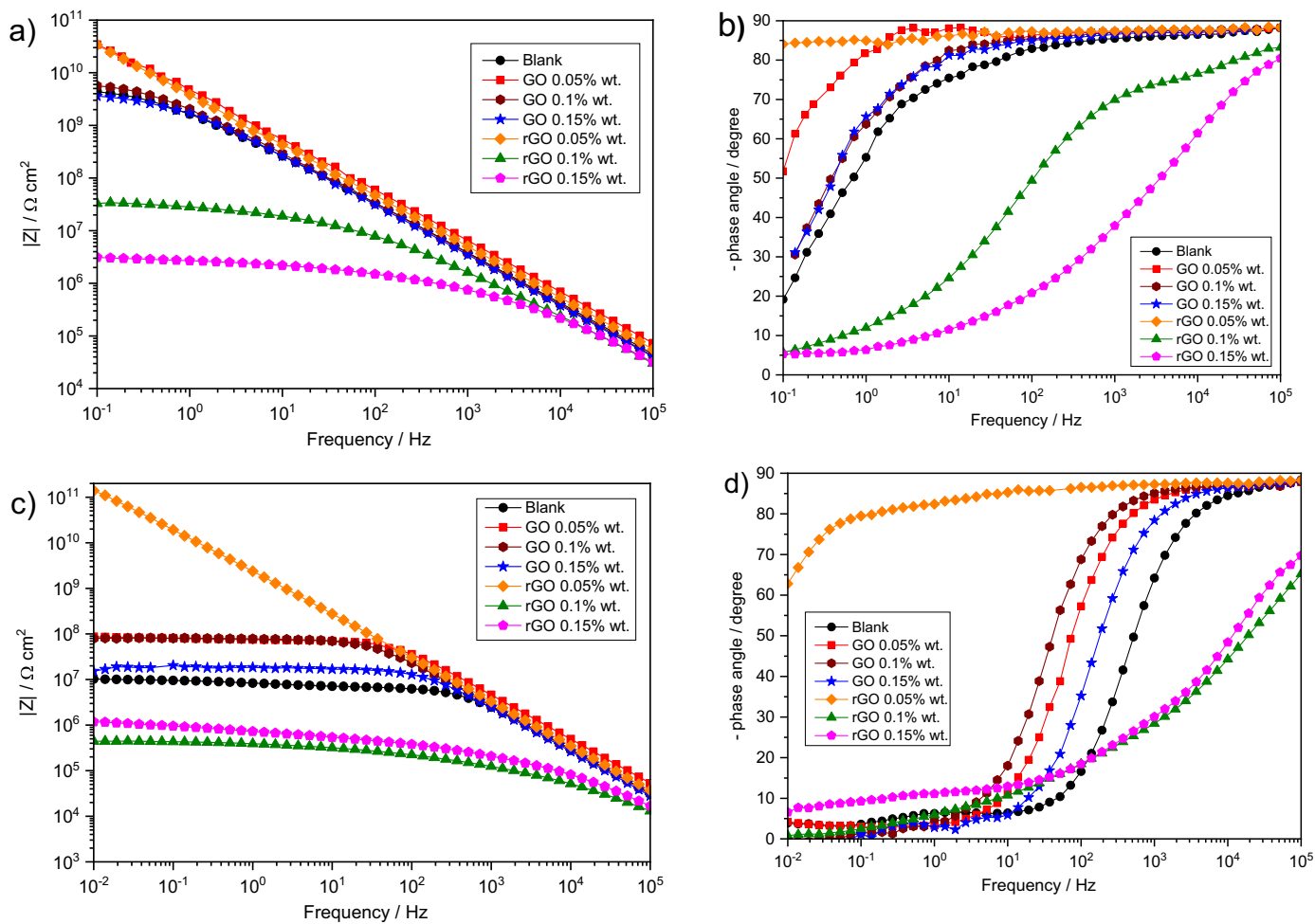

**Fig. 10.** Bode plots of galvanized steel coated with the different formulations after 1 h (a, b) and 17 days (c, d) of immersion in 0.1 M NaCl solution.

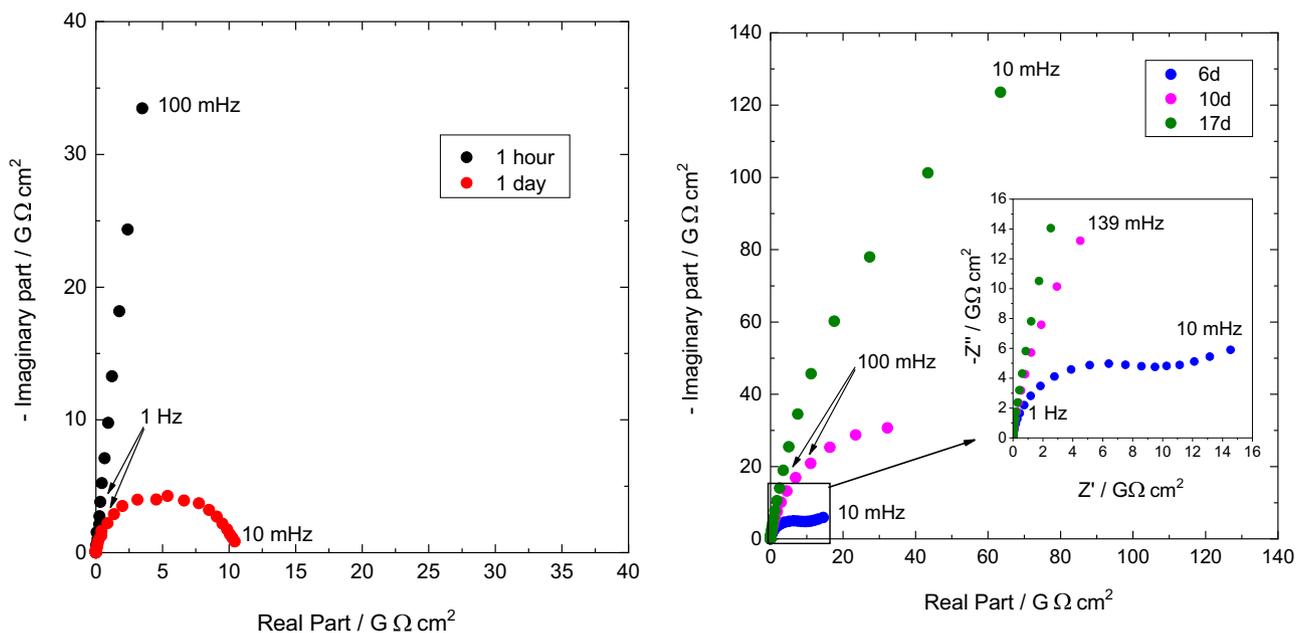

**Fig. 11.** Impedance evolution of rGO 0.05 wt% system with immersion time in 0.1 M NaCl solution.

a)

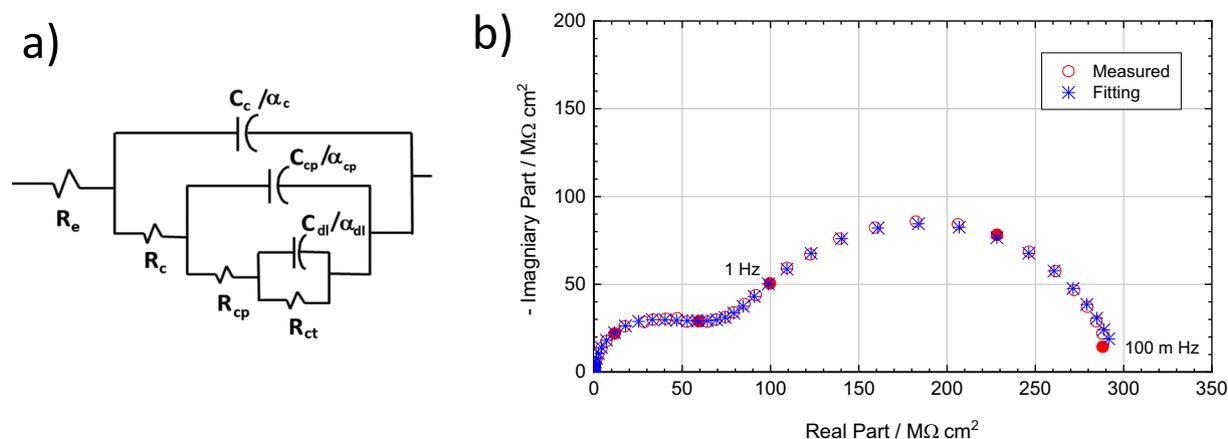

b)

**Fig. 12.** EEC used to model the behaviour of coated galvanized steel samples (a). Experimental (O) and fitted (S) data using the EEC showed in Fig. 12a (b), the best fitting parameters obtained are: $R_e = 1.3 \times 10^{-3} \ \Omega \ cm^2$, $R_c = 5.9 \times 10^7 \ \Omega \ cm^2$; $C_c = 5.5 \times 10^{-11} \ F \ cm^{-2}$, $R_{cp} = 3.3 \times 10^7 \ \Omega \ cm^2$; $C_{cp} = 5.5 \times 10^{-10} \ F \ cm^{-2}$, $R_{ct} = 2.1 \times 10^8 \ \Omega \ cm^2$; $C_{dl} = 3.4 \times 10^{-9} \ F \ cm^{-2}$.

the total resistance of the film, even if there is no electronic contact between them.

### 3.4. Anticorrosive properties of the coatings

The protective properties of the films applied on galvanized steel substrates were studied by EIS measurements in 0.1 M NaCl solution. The impedance evolution with immersion time is shown in Fig. 10.

One hour of immersion is enough to distinguishing different performances depending on the system (Fig. 10a and b). It is noticeable the low impedance values that were observed for samples with rGO 0.1 wt% and 0.15 wt%. However, considering the behaviour observed in the free film experiments, the low impedance values recorded could be expected and corroborate the presence of percolating pores. Additionally, the aggregation may facilitate the contact between the conductive rGO particles and the metallic substrate in the defective areas, expanding the rGO/metallic substrate micro-galvanic couple. Therefore the protection properties would be lost [12,39].

The other systems exhibit good barrier features, with impedance values higher than $10^9 \ \Omega \ cm^2$, in good agreement with the capacitive behaviour observed in the free film experiments. We can also observe that the better behaviour corresponds to the films doped with 0.05 wt% of GO and rGO, which suggests that small amounts of these pigments can be well-dispersed into the resin, improving its barrier properties and counteracting the harmful effect of the porosity generated during the reticulation of the resin.

The differences are more evident with longer immersion times (Fig. 10c and d). Most of the formulations decrease their protective properties as the lower impedance values reflect. Even though, systems doped with GO maintain higher impedance values than unpigmented resin, which states the benefit provided by the GO.

The exception is the rGO 0.05 wt% formulation, where the impedance after 17 days of immersion is even higher than at the beginning. This increase should be associated with the presence of a protective mechanism added to the barrier one. Fig. 11 can help to explain. In the first 24 h of immersion, the electrolyte uptake is fast (see also Fig. 7), so it gets in contact with the rGO particles homogeneously distributed into the resin. These particles provide an additional cathodic surface, accelerating the corrosion process, and the impedance decreases drastically. As the corrosion progresses, a uniform film of zinc corrosion products is generated that passivates the surface, which translates into an impedance increase. The identification of the generated corrosion products is analysed in the 3.5 section.

The Electrical Equivalent Circuit (EEC) employed to model the observed behaviour is displayed in Fig. 12a, and consists of three-time constants hierarchically distributed. $R_e$ denotes de electrolyte resistance, the high-frequency time constant, $R_cC_c$, accounts for the dielectric properties of the coating, at medium frequency, an additional time constant, $R_{cp}C_{cp}$, can be observed, and it is related to the corrosion products, and the low-frequency time constant, $R_pC_p$, is associated with the corrosion process. All the time constants are affected by the Cole-Cole parameter dispersion, $\alpha_i$, which meaning has above described. This EEC is commonly used to model the behaviour of coated metallic substrates [40–42]. We can check the quality of the fitting in Fig. 12b.

The evolution of the parameters associated with the high-frequency time constant, $R_c$ and $C_c$, is shown in Fig. 13a. The coating resistance tends to decrease with immersion time, whereas the capacitance experiences an increase mainly during the first days of immersion. This evolution is expected and is the consequence of the electrolyte entrance, which is more pronounced for rGO 0.1 wt% and rGO 0.15 wt% formulations. These results are in good agreement with those obtained from the free films. The second time constant, $R_{cp}$ and $C_{cp}$, located in the middle frequency range, experienced a similar evolution. However, the fluctuations are higher (see Fig. 12b), suggesting the dynamic nature of the generation/dissolution of the zinc corrosion products. This time constant was not observed in the rGO 0.05 wt% system; its contribution may overlap with the film response. The evolution of the low-frequency time constant, $R_{ct}$ and $C_{dl}$, is depicted in Fig. 13c; as a general tendency, the charge transfer resistance decreases, and the double layer capacitance increases with immersion time, typical of the corrosion development. The rGO 0.05 wt% system follows a distinctive way, after an initial activation period, characterised by a sharp resistance decrease and a capacitance increase; the charge transfer resistance markedly increases and the capacitance decreases, which confirms the passivation of the metallic surface.

### 3.5. Characterization of the corrosion products

Fig. 14 depicts the surface state of the samples at the end of the immersion test. The degradation is evident in GO 0.15 wt%, rGO 0.1 wt% and rGO 0.15 wt% specimens, according to the impedance results.

Although the presence of the corrosion products is visible in several samples, their quantity is not enough to characterize them. For this reason, some specimens were kept in the same solution for an extended time until the amount was enough to perform their characterization by XRD and SEM/EDX techniques. Fig. 15 summarises the results.

Fig. 15a clearly shows the non-uniformity of zinc corrosion via the aspect of a GO 0.15 wt% sample after 60 days of immersion in the 0.1 M NaCl solution. Two types of blisters are clearly distinguished by their colours: black and white, suggesting the generation of corrosion products of different nature. A dark area surrounding both is also visible, illustrating the delamination spread. The SEM images displayed in Fig. 15b-c discloses two distinct morphologies depending on the kind of blisters. Thus, inside black blisters, it may appreciate corrosion products with flake-like morphology, whereas the corrosion products located in the white blisters have a globular-shape morphology. The EDX spectra (see Fig. 15d) reveal that the morphology and chemical nature of the generated corrosion products changes according to the type of blisters. The semiquantitative results extracted from the EDX measurements were included in Fig. 15d. The higher peak intensity corresponds to zinc and oxygen elements related to zinc corrosion products. The presence of carbon should be associated with the polymer since the generation of zinc carbonates is not probable [43] under the coating. Additional chloride signal was detected only in the black blisters indicating the presence of a kind of hydroxyl chloride [44]. XRD measurements were performed to identify these corrosion products; the results are shown in Fig. 16.

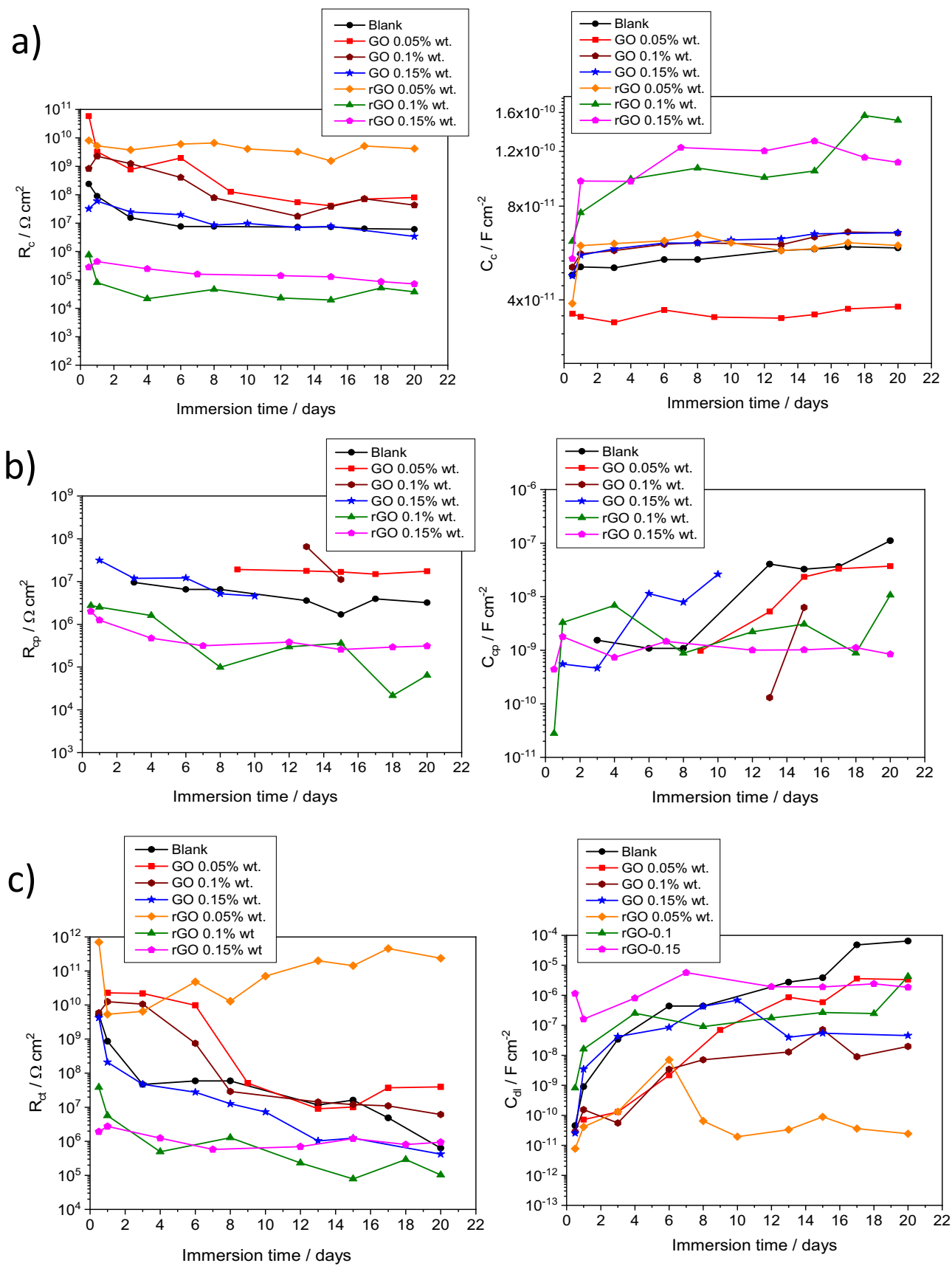

**Fig. 13.** Evolution of the fitting parameters corresponding to the dielectric properties of the films ($R_pC_p$) (a), the corrosion products ($R_{cp}C_{cp}$) (b), and the corrosion process ($R_{ct}C_{dl}$) (c) for the studied systems immersed in 0.1 M NaCl solution during 20 days.

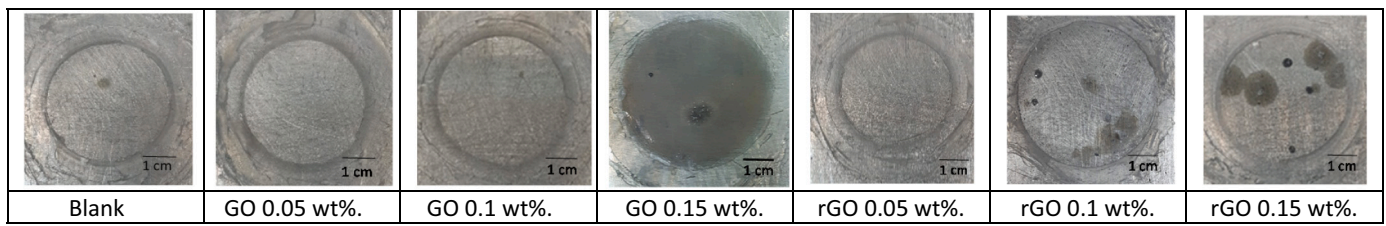

| Blank | GO 0.05 wt%. | GO 0.1 wt%. | GO 0.15 wt%. | rGO 0.05 wt%. | rGO 0.1 wt%. | rGO 0.15 wt%. |

**Fig. 14.** Photographs of the samples showing the aspects at the end of the immersion test.

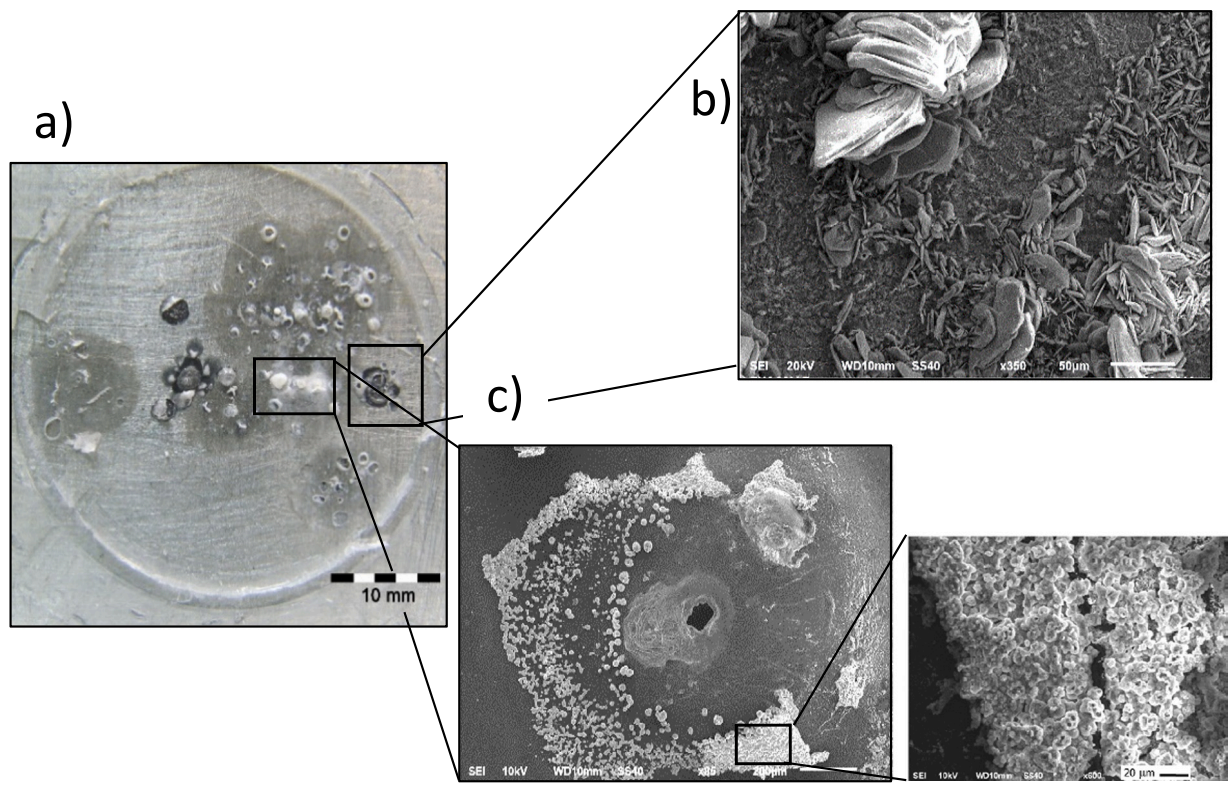

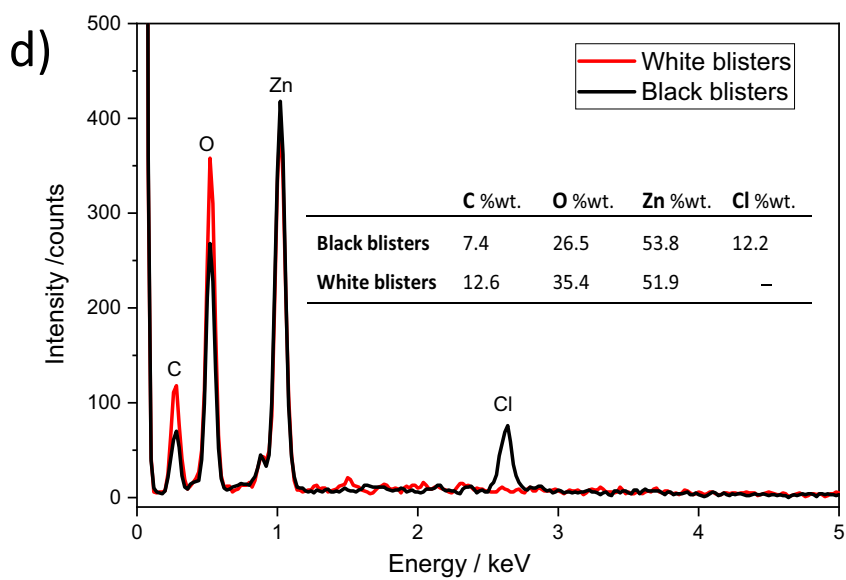

**Fig. 15.** Picture of GO 0.15 wt% system after 60 days of immersion in 0.1 M NaCl solution (a), SEM images corresponding to the corrosion products inside the black (b) and white (c) blisters, including in this last a detail of the selected area, and the EDX spectra of both types of blisters, the quantitative analysis is inserted (d).

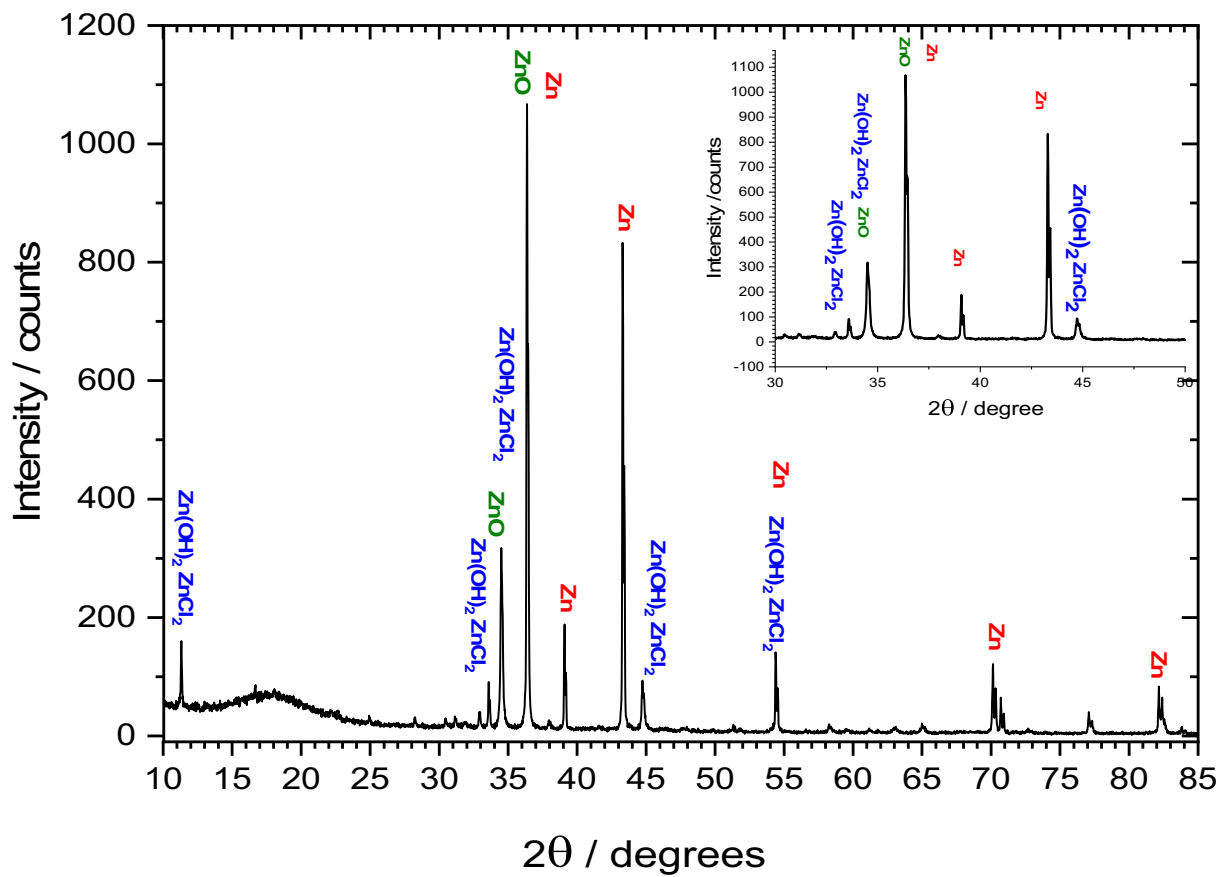

**Fig. 16.** XRD pattern of GO 0.15 wt%. system after 60 days of immersion in 0.1 M NaCl solution.

Two corrosion products were identified: zincite, ZnO; and a zinc hydroxychloride, $Zn(OH)_2ZnCl_2$. The absence of zinc carbonate species corroborates the interpretation given to the carbon peak observed in the EDX spectra. On the other hand, the presence of ZnO confirms the barrier effect of the polymer to the carbon dioxide entrance. This fact has already been stated by M.C. Bernard et al. [44,45] and in previous studies of our group [43], where we demonstrated that zincite effectively passivates the zinc surface.

The above conclusion may explain the excellent behaviour observed in the rGO 0.05 wt% formulation. The low amount of rGO promotes a uniform distribution into the resin and, probably, with a minor network disruption. When the water comes through the film, the corrosion process initiates fast, generating only ZnO because the chloride ions take longer to diffuse through the film. The result will be the generation of a

ZnO film that passives to the galvanized steel.

## 4. Conclusions

The present study focuses on the GO/rGO additions to a water-borne resin. From the results presented, we drew the following main conclusions:

The zeta potential values confirmed the stability of the GO and rGO aqueous suspensions. As the measurements were performed at the pH of the water-borne resin, good dispersibility is guaranteed, at least at low concentrations. The 3D surface profiles showed certain aggregation for the higher concentrations, mainly rGO 0.1 wt%, rGO 0.15 wt%, and GO 0.15 wt%.

The transport properties of the free films clearly show the consequences of aggregation. Thus, the rGO 0.1 wt% and rGO 0.15 wt% formulations developed percolating pores after a few hours in contact with the electrolyte. That negative effect was also corroborated by monitoring the relative variation of the real part of the permittivity ($\Delta\varepsilon'$) with the immersion time, where the higher increase corresponded to these two formulations. Regardless of that, all the systems followed a similar pattern: an initial fast $\Delta\varepsilon'$ increase, followed by a slower $\Delta\varepsilon'$ rise. This behaviour was explained considering an initial fast electrolyte uptake at the more accessible areas of the films, and a slower process due to the water interaction with the bulk matrix or the diffusion through the percolating pores.

The study of the applied films in dry conditions showed that the complex permittivity followed the Cole-Cole type distribution due to the heterogeneous nature of the films with pores or flaws. This effect was more substantial in the doped films. In addition, the presence of particles disrupted the polymer crosslinking.

EIS allowed assessing the anticorrosive properties. The results corroborated the poor performance of the films doped with rGO 0.1 wt% and rGO 0.15 wt%. The explanation considered two factors: the high porosity of these films, and the establishment of micro galvanic coupling between the rGO particles and the metallic substrate. The other formulations provided better barrier properties than the undoped film. The rGO 0.05 wt% formulation exhibited good long-term performance justified by two synergistic effects: the barrier mechanism and the zinc surface passivation by the generation of the ZnO as the primary corrosion product in these conditions.

## Declaration of competing interest

The authors declare no conflict of interest.

## Data availability

Data will be made available on request.

## Acknowledgments

Akzo Nobel Coating S.L. is acknowledged for supplying the resin.